\documentclass[sigconf]{acmart}

\setcopyright{none}              
\acmConference[]{}               
\acmDOI{}                        
\acmISBN{}                       
\renewcommand\footnotetextcopyrightpermission[1]{} 
\makeatletter
\let\@authorsaddresses\@empty    
\makeatother
\AtBeginDocument{%
  }

\usepackage{xcolor}
\usepackage{mathtools}
\usepackage{amsthm}
\newtheorem{definition}{Definition}
\newtheorem{corollary}{Corollary}
\newtheorem{remark}{Remark}
\newcommand{\F}{\mathbb{F}}
\newcommand{\R}{\mathbb{R}}
\newcommand{\E}{\mathbb{E}}

\begin{document}

\title{Feedback Lunch: Learned Feedback Codes for Secure Communications}




\author{Yingyao Zhou}
\orcid{0009-0000-9675-439X}
\affiliation{%
  \institution{University of Illinois Chicago}
  \city{Chicago}
  \state{IL}
  \country{USA}}
\email{yzhou238@uic.edu}

\author{Natasha Devroye}
\orcid{0000-0002-1619-4095}  
\affiliation{%
  \institution{University of Illinois Chicago}
  \city{Chicago}
  \state{IL}
  \country{USA}}
\email{devroye@uic.edu}

\author{Onur G\"{u}nl\"{u}}
\orcid{0000-0002-0313-7788}
\affiliation{%
  \institution{Technische Universität Dortmund}
  \city{Dortmund}
  \state{North Rhine-Westphalia}
  \country{Germany}
}
\affiliation{%
  \institution{Linköping University}
  \city{Linköping}
  \state{Östergötland}
  \country{Sweden}
}
\email{onur.guenlue@tu-dortmund.de}


\begin{abstract}
  We consider reversely-degraded secure-communication channels, for which the secrecy capacity is zero if there is no channel feedback. Specifically, we focus on a seeded modular code design for the block-fading Gaussian wiretap channel with channel-output feedback, combining universal hash functions for security and learned feedback-based codes for reliability. The trade-off between communication reliability and information leakage is studied, illustrating that feedback enables agreeing on a secret key shared between legitimate parties, overcoming the security advantage of the eavesdropper. Our findings motivate code designs for sensing-assisted secure communications in the context of integrated sensing and communication (ISAC).
\end{abstract}





\maketitle

\section{Introduction}
Secure communication can be achieved through key-based cryptography, which relies on securely shared secret keys, and physical-layer security (PLS), which leverages the physical properties of the channel to ensure data confidentiality~\cite{bloch2011physical, OurJSAITtutorial}. Wyner introduced the wiretap channel (WTC), a fundamental model in PLS that enables reliable and secure communication when the eavesdropper's channel is a degraded version of the legitimate receiver's channel, resulting in a positive secrecy capacity~\cite{wyner1975wire}. Although feedback does not increase the channel capacity of memoryless channels, it improves the secrecy capacity \cite{ahlswede2006transmission, oursecureISACJSAIT}. Here, we focus on the most challenging wiretap channel case, where the eavesdropper has a statistical advantage over the legitimate receiver, i.e., a reversely-degraded wiretap channel with output feedback (RD-WTC-F). We design neural feedback codes for this setting.

The modular coding scheme in \cite{bellare2012cryptographic} treats the security and reliability separately. Several works~\cite{rana2023short,seifert2024deep} build on this idea for the Gaussian WTC without feedback. However, feedback (which can be obtained, for instance, through reflections) is an intrinsic component of, e.g., mono-static integrated sensing and communication (ISAC) applications \cite{VDEJCASPositionPaper,oursecureISACJSAIT}. Recently, deep-learned error-correcting feedback codes (DL-ECFCs) have been developed, as in \cite{ankireddy2024lightcode}, for the additive white Gaussian noise channel with output feedback (AWGN-F), outperforming analytical linear codes in most cases.

\paragraph{Main Contributions}
We extend the modular coding design \cite{bellare2012cryptographic} by combining universal hash functions with the current state-of-the-art DL-ECFC, Lightcode~\cite{ankireddy2024lightcode}, to consider block-fading Gaussian WTC-F. We design and compare multiple feedback codes for the challenging case where the eavesdropper experiences the same or lower noise variance than the legitimate receiver (i.e. RD setting). Non-feedback codes fail to achieve a positive secrecy rate in this setting. In contrast, our feedback coding scheme obtains high reliability for positive secrecy rates at non-asymptotic blocklengths by leveraging feedback to generate secret keys shared between legitimate parties. Even with noisy feedback, legitimate parties achieve a security advantage over the eavesdropper. We refer to these feedback gains as a ``feedback lunch'', akin to a somewhat ``free lunch'' rather than the ``no-free-lunch'' theorems in security applications~\cite{zhang2022no}. To further enhance the secrecy performance, we introduce a new loss function incorporating an information-leakage constraint, enabling the learned model to reduce leakage while maintaining high reliability. 

\paragraph{Notation}
Random variables are denoted by capital letters, specific realizations by lowercase letters, and vectors in boldface. Define the binary field $\F_2 = \{0,1\}$, $n$-dimensional real vectors $\R^n$, and positive reals $\R_{+}$. The sequence $[a\!:\!b\!:\!c]$ denotes $[a, a+b, \ldots, c]$. Subscripts denote time indices, and superscripts indicate vector length (e.g., $\mathbf{Y}^n = (Y_1, \ldots, Y_n)$). Set cardinality is $\mathrm{card}(\mathcal{X})$, and $[x]^+ = \max\{x, 0\}$. The standard Gaussian complementary cumulative distribution function is denoted as $Q(x)$.  Specifically, $k$ is the message length, $q$ the security-layer output length, and $n$ the blocklength.

\section{System Model}
\begin{figure}[!t]
    \centering
\includegraphics[width=0.98\columnwidth]{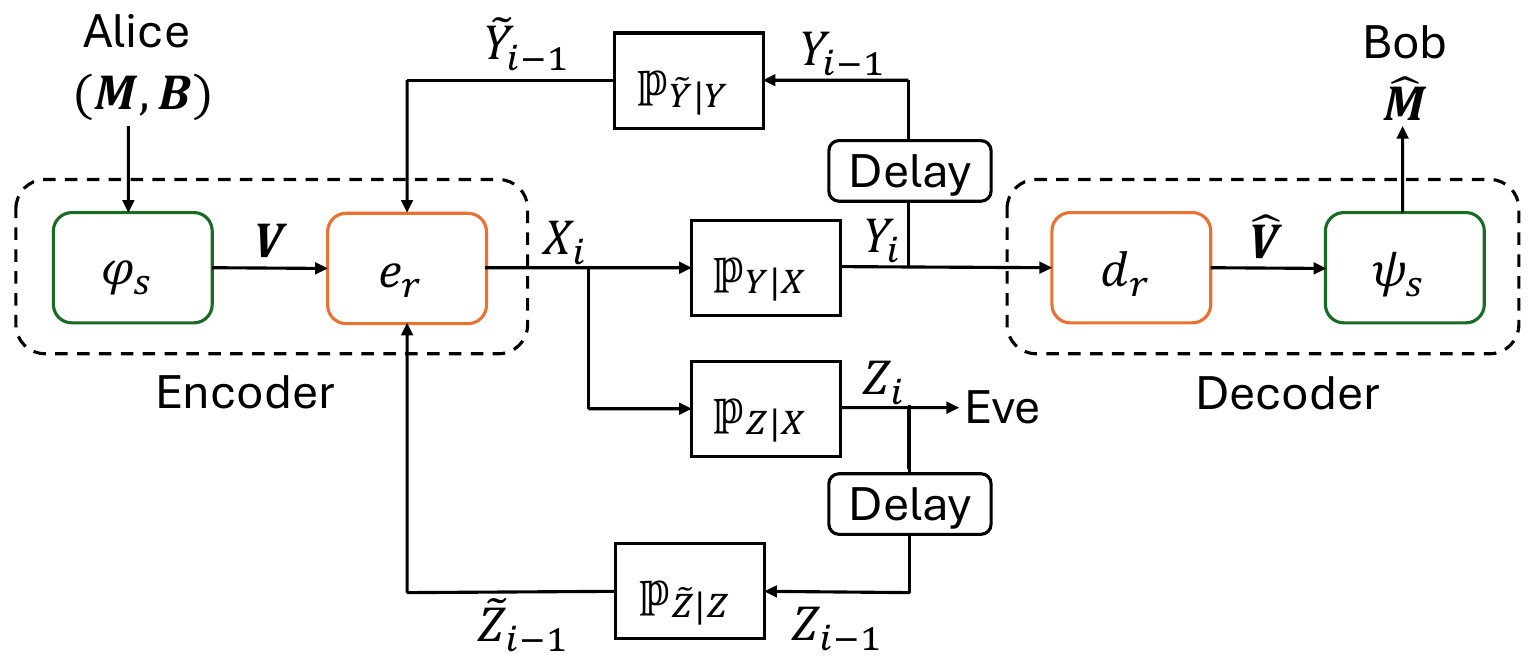}
 \Description{A block diagram illustrating a secure communication system where Alice encodes the message, Bob decodes it, and Eve attempts to intercept it, with output feedback from both Bob and Eve to Alice. The encoder and decoder consist of a security layer and a reliability layer.}
    \caption{Design of modular deep-learned feedback wiretap codes. The security and reliability layers are given by $(\varphi_s, \psi_s)$ and $(e_r, d_r)$, respectively.}
    \label{fig:wiretap}
\end{figure}

 Consider first a real-valued block-fading RD-WTC-F with $\mathbb{P}(Y,Z|X)$, as depicted in Figure~\ref{fig:wiretap}.
 The transmitter, Alice, sends a message $\mathbf{M} \in \F_2^{k}$ to the legitimate receiver, Bob, over $n$ channel uses, which are also observed by the eavesdropper, Eve. The code rate is defined as $R = {k}/{n}$. At each channel use $i \in [1\!:\!1\!:\!n]$, the real-valued forward fading channels to Bob and Eve are modeled as
\begin{align}
Y_i = H_{Y}X_i + N_{Y,i},\qquad Z_i = H_{Z}X_i + N_{Z,i}
\end{align}
where $N_{Y,i} \sim \mathcal{N}(0,\sigma_{N_Y}^2)$ and $N_{Z,i} \sim \mathcal{N}(0,\sigma_{N_Z}^2)$ denote independent and identically distributed (i.i.d.) Gaussian noise components. The fading coefficients $H_Y$ and $H_Z$ remain constant within each block of length $n$, and are i.i.d. across different blocks according to the Rayleigh probability density functions
\begin{equation}
    f_{H}(h) = \frac{2h}{\sigma_{H}^2}\exp\left(-\frac{h^2}{\sigma_{H}^2}\right),\qquad  h\geq 0
\end{equation}
where $\mathbb{E}[H^2] = \sigma_H^2$. Suppose $H_Y \sim \mathrm{Rayleigh}(\sigma_{H_Y})$ and $H_Z \sim \mathrm{Rayleigh}(\sigma_{H_Z})$ are independent of each other and independent of the additive Gaussian noise components. The transmitted symbols $X_i\in \R$ are subject to the average power constraint $\frac{1}{n} \mathbb{E}\left(\sum_{i = 1}^nX_i^2\right) \leq P$. We consider real-valued symbols to ensure differentiability during neural network training, and extensions to discrete constellations via quantization are left for future work, for which the straight-through estimator can be used to enable training with discrete symbols. Assume further that there is channel-output feedback available at the transmitter, which is delayed, possibly noisy, and from both Bob and Eve. Such feedback could be obtained via reflections from both Bob and Eve, which does not necessarily require an active transmission from them \cite{VDEJCASPositionPaper}. Feedback can be either noiseless or noisy, with the noisy case modeled as $\tilde{Y}_{i-1} = Y_{i-1} + \tilde{N}_{Y,i-1}$ and $\tilde{Z}_{i-1} = Z_{i-1} + \tilde{N}_{Z,i-1}$, where $\tilde{N}_{Y,i-1} \sim \mathcal{N}(0,\tilde{\sigma}_{N_Y}^2)$ and $\tilde{N}_{Z,i-1} \sim \mathcal{N}(0,\tilde{\sigma}_{N_Z}^2)$. For Bob, we define the average forward signal-to-noise ratio (SNR) as $\mathrm{\bar{S}}_{Y,f} = \sigma_{H_Y}^2 P/\sigma_{N_Y}^2$ and the feedback SNR as $\mathrm{S}_{Y,fb} = P/\tilde{\sigma}_{N_Y}^2$. The corresponding SNRs for Eve, $\mathrm{\bar{S}}_{Z,f}$ and $\mathrm{S}_{Z,fb}$ are defined analogously.

As in Figure~\ref{fig:wiretap}, consider the modular coding scheme proposed in~\cite{bellare2012cryptographic}, which consists of a security layer $(\varphi_s, \psi_s)$ and a reliability layer $(e_r, d_r)$. Separating these two layers facilitates independent control of information leakage and reliability. The security layer aims to control the leakage of the message $\mathbf{M}$ to Eve through her observations $\mathbf{Z}^n$. The security encoder uses a uniformly distributed random bit sequence $\mathbf{B} \in \F_2^{q-k}$ to produce $\mathbf{V} = \varphi_s(\mathbf{M}, \mathbf{B}) \in \F_2^{q}$. The security decoder recovers the message from the output of the reliability decoder via $\widehat{\mathbf{M}} = \psi_s(\widehat{\mathbf{V}})$. Moreover, the reliability encoder $e_r$ generates the transmitted symbol $X_i$ using the output of the security encoder $\mathbf{V}$ and past feedback from Bob and Eve as
\begin{equation}
    X_i = {e_r}(\mathbf{V}, \mathbf{X}^{i-1}, \tilde{\mathbf{Y}}^{i-1}, \tilde{\mathbf{Z}}^{i-1}), \quad  \forall i \in [1\!:\!1\!:\!n].
\label{eq:er}
\end{equation}
After $n$ channel uses, the reliability decoder $d_r$ estimates $\mathbf{V}$ from the received noisy codewords. In the fading case, we assume that perfect channel state information (CSI) is available at the receivers. Formally,
\begin{equation}
\widehat{\mathbf{V}} =
\begin{cases}
d_r(Y_1, \ldots, Y_n, H_Y), & \text{fading}, \\
d_r(Y_1, \ldots, Y_n), & \text{non-fading}.
\end{cases}
\end{equation}
We denote the effective code rate of the reliability layer as $R_r = {q}/{n}$. The functions $e_r$ and $d_r$ represent learned models.

Codes for the wiretap channel aim to ensure reliable transmission to Bob while maintaining the confidentiality of the message. To quantify information leakage, we define the security metric as $L_{\text{eve}} \coloneqq I(\mathbf{M}; \mathbf{Z}^n\mid H_Z)$ (or $I(\mathbf{M}; \mathbf{Z}^n)$ for the non-fading case), representing the mutual information (MI) between the message and the observations at Eve, measured in bits. Similarly, we define $I_{\mathrm{bob}} \!\coloneqq\! I(\mathbf{M}; \mathbf{Y}^n\mid H_Y)$ (or $I(\mathbf{M}; \mathbf{Y}^n)$ in the non-fading case). For reliability, we consider the block error rate (BLER) at Bob, defined as $\mathrm{BLER} \coloneqq \mathbb{P}(\mathbf{M} \neq \widehat{\mathbf{M}})$. To establish a performance baseline, we first consider the non-fading scenario where $H_Y = H_Z = 1$. In this case, the secrecy capacity is the maximum rate at which information can be transmitted both reliably and securely over a wiretap channel in the asymptotic regime, {as $n\rightarrow\infty$}. For the Gaussian wiretap channel without feedback, the secrecy capacity is given by~\cite{leung1976multi}
\begin{equation}
    \frac{1}{2}\left[\log\left(1+P/\sigma_{N_Y}^2\right) - \log\left(1 + P/\sigma_{N_Z}^2\right)\right]^+.
\label{eq:nofb}
\end{equation}
Thus, the secrecy capacity is zero if $\sigma_{N_Y}^2\geq \sigma_{N_Z}^2$, as in a RD-WTC. We next provide the secrecy capacity with noiseless feedback.
\begin{corollary}[\cite{ahlswede2006transmission}] 
For a RD wiretap channel with $\mathbb{P}(Y, Z|X) = \mathbb{P}(Z|X)\mathbb{P}(Y|Z)$, the secrecy capacity with noiseless feedback is\footnote{Although stated for discrete random variables, this result is shown to extend to continuous-alphabet cases, such as the Gaussian wiretap channel.}  
\begin{equation}
    \max_{\mathbb{P}_{X}} \min[H(Y|Z), I(X;Y)].
\end{equation}
\end{corollary}
Thus, with feedback, the secrecy capacity can become positive, unlike the non-feedback secrecy capacity in~\eqref{eq:nofb}, which is zero in this regime.  We demonstrate this feedback lunch property by constructing practical finite-length feedback codes.

\section{Prior Work}
In this section, we review two feedback coding schemes for the AWGN channel with \textit{noiseless} feedback: the Schalkwijk–Kailath (SK) scheme~\cite{schalkwijk1966coding} and the POWERBLAST (PB) scheme~\cite{ankireddy2024lightcode}.
For the WTC-F, we assume that feedback is available only from Bob. These schemes, therefore, serve as the reliability layer in our framework. When combined with the security layer, the resulting schemes are denoted as WTC-SK and WTC-PB, respectively.

The SK scheme asymptotically achieves the AWGN-F channel capacity $\frac{1}{2}\log(1 + P/\sigma_{N_Y}^2)$~\cite{schalkwijk1966coding}. It has also been used to achieve secrecy-capacity results for certain Gaussian WTC-F settings~\cite{wei2019some}. In the first round, $\mathbf{M}\in \F_2^k$ ({corresponding to $\mathbf{V}\in \F_2^q$ in WTC-SK}) is mapped to a pulse amplitude modulation (PAM) symbol $\Theta \in \{\pm{\eta}, \pm{3\eta}, \dots, \pm{(2^{k}-1)\eta}\}$, where $\eta = \sqrt{\frac{3}{2^{2k}-1}}$ and the transmitted signal is $X_1 = \sqrt{P}\Theta$. In the  subsequent rounds, let $\epsilon_i = \widehat{\Theta}_i - \Theta$ with mean squared error $D_i = \mathbb{E}(\epsilon_i^2)$, where $\widehat{\Theta}_i$ is the estimated symbol at time $i$. The transmitter sends $X_i = \sqrt{\frac{P}{D_{i-1}}}\epsilon_{i-1}$. At the receiver, the linear minimum mean square error estimate is $\widehat{\Theta}_1= \frac{\sqrt{P}}{P + \sigma_{N_Y}^2}Y_1$. For $i \geq 2$, the error is estimated as $\hat{\epsilon}_{i-1} = \frac{\sqrt{PD_{i-1}}}{P + \sigma_{N_Y}^2} Y_i$ and the estimate is updated as $\widehat{\Theta}_i = \widehat{\Theta}_{i-1} - \hat{\epsilon}_{i-1}$. Finally, the receiver decodes by mapping $\widehat{\Theta}_n$ to the nearest PAM symbol. The SK scheme achieves doubly exponential error decay in $n$ in terms of BLER. The PB~\cite{ankireddy2024lightcode} scheme is a variation of SK that differs only in the final round, where the transmitter sends a discrete symbol for the estimated PAM index error $X_n = \sqrt{\frac{P}{D_I}}(I(\widehat{\Theta}_{n-1}) - I(\Theta))$ with $D_I = \mathbb{E}[(I(\widehat{\Theta}_{n-1}) - I(\Theta))^2]$ and $I(\Theta)\in [1\!:\!1\!:\!2^{k}]$. At the receiver, a maximum a posteriori decoder is used to recover the PAM symbol.

\section{Proposed Coding Schemes}\label{sec:coding}
We next propose practical feedback coding schemes for WTC-F by extending the modular coding scheme. To measure their security performance, we also estimate the information leakage using neural networks, which is later used to improve their overall performance.

\subsection{Security Layer}
To limit the information leakage, we use 2-universal hash functions (2-UHF) $\varphi_s$ and their inverses $\psi_s$, as defined below.

\begin{definition}[2-UHF~\cite{carter1977universal}]\label{def:2uhf}
 Let $\mathcal{X}$ and $\mathcal{Y}$ be finite sets. A family of functions $\mathcal{H}$, where each $H: \mathcal{X} \mapsto \mathcal{Y}$, is 2-universal if $\forall x_1, x_2 \in \mathcal{X}$ with $x_1\neq x_2$, we have
\begin{equation}
    \mathbb{P}_{H\sim\mathcal{H}}(H(x_1) = H(x_2)) \leq \text{card}(\mathcal{Y})^{-1}
\end{equation}
where $H$ is chosen uniformly at random from $\mathcal{H}$. 
\end{definition}

Let the seed space be $\mathcal{S} \coloneqq \F_2^q \backslash \{\mathbf{0}\}$, where the seed $\mathbf{S} \in \mathcal{S}$ is shared among the legitimate parties. For each message $\mathbf{M} \in \F_2^{k}$, Alice samples a uniformly distributed random bit sequence $\mathbf{B} \in \F_2^{q-k}$. The security encoder $\varphi_s$ is defined as
\begin{equation}
\label{eq: security_encoder}
    \varphi_s: (\mathbf{M}, \mathbf{B}) \mapsto \mathbf{S}^{-1} \odot (\mathbf{M} \| \mathbf{B})
\end{equation}
where $(\cdot\|\cdot)$ denotes the concatenation of two sequences, and $\odot$ represents multiplication in $\mathrm{GF}(2^q)$. The corresponding security decoder recovers the estimated message $\widehat{\mathbf{M}}$ using
\begin{equation}
\label{eq: security_decoder}
    \psi_s: \widehat{\mathbf{V}}  \mapsto (\mathbf{S} \odot  \widehat{\mathbf{V}})_k
\end{equation}
where $(\cdot)_k$ denotes selection of the $k$ most significant bits.

\subsection{Reliability Layer}
\begin{figure}[!t]
    \centering
\includegraphics[width=0.95\columnwidth]{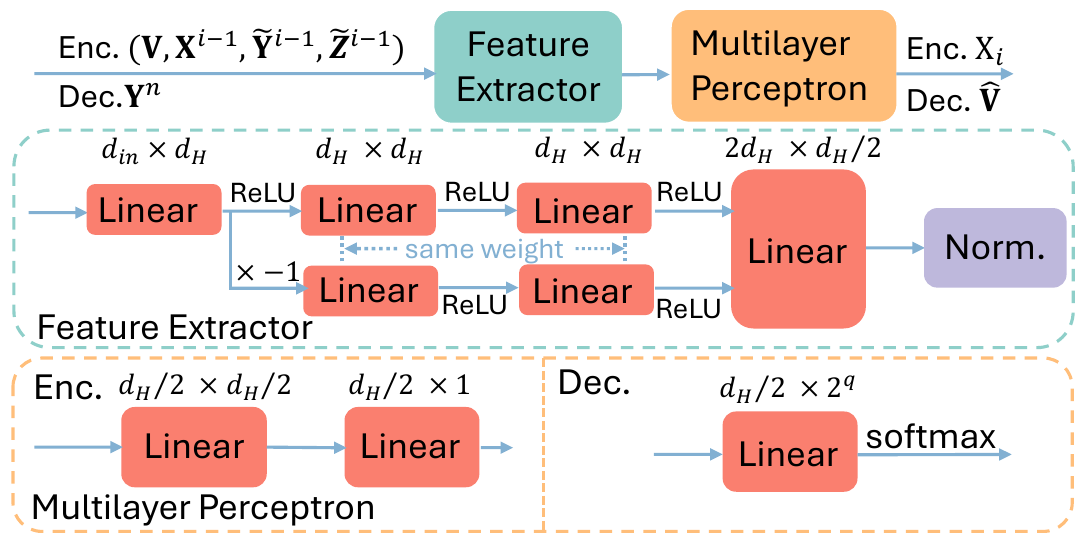}
\Description{The structure of the reliability layer, where both the reliability encoder and decoder consist of a feature extractor and a multilayer perceptron.}
    \caption{Detailed structure of the reliability layer.}
    \label{fig:lightcode}
\end{figure}
The reliability layer $(e_r, d_r)$ can be implemented using neural networks. Unlike the original Lightcode~\cite{ankireddy2024lightcode}, WTC-Lightcode incorporates feedback from both Bob and Eve, as in the broadcast case~\cite{zhou2025learned} (Figure~\ref{fig:wiretap}), and augments Lightcode with the security layer $(\varphi_s, \psi_s)$. Both $e_r$ and $d_r$ consist of a feature extractor and a multi-layer perceptron, as shown in Figure~\ref{fig:lightcode}, where $d_H=32$. As $n$ increases, the parameter size grows rapidly, making training difficult, so we focus on short blocklengths, with longer transmissions partitioned into smaller chunks~\cite{ankireddy2024lightcode}.
The reliability encoder and decoder are trained jointly to minimize the categorical cross entropy (CCE)
\begin{align}
    J_{\text{CCE}} = -\frac{1}{B}\sum_{i=1}^B\left(\sum_{j=1}^{C_V}p_{ij}\log(\hat{p}_{ij})\right)
\end{align}
where $B$ is the batch size, $C_V = 2^q$, $p_{ij} = 1$ if class $j$ is correct for instance $i$ and 0 otherwise, and $\hat{p}_{ij} \in \mathbb{R}$ is the predicted probability.

\subsection{Information Leakage}
The joint or marginal distributions are often unknown in closed form, making direct computation of MI infeasible. We therefore approximate $I_{\text{bob}}$ using Mutual Information Neural Estimator (MINE) \cite{belghazi2018mutual} and  $L_{\text{eve}}$ with $f$-divergence discriminative mutual information estimators ($f$-DIME)~\cite{letizia2024fDIME}.

MINE uses the dual representation of the Kullback-Leibler (KL) divergence, yielding the lower bound
\begin{equation}
   {I}_{\text{bob}}  \approx  \sup_{\omega \in \Omega}\bigg[\E_{\mathbb{P}_{\mathbf{M}\mathbf{Y}^n}}[T_{\omega}(\mathbf{m},\mathbf{y}^n)] - \log\!\left(\E_{\mathbb{P}_{\mathbf{M}}\mathbb{P}_{\mathbf{Y}^n}}[e^{T_{\omega}(\bar{\mathbf{m}},\bar{\mathbf{y}}^n)}]\right)\bigg]
\end{equation}
where $T_{\omega}:\F_2^k\times\R^n \rightarrow \R$ is a neural network-parameterized function. Here, $(\mathbf{m}, \mathbf{y}^n)$ are samples drawn from the joint distribution $\mathbb{P}_{\mathbf{M}\mathbf{Y}^n}$, while $(\bar{\mathbf{m}}, \bar{\mathbf{y}}^n)$ are independent samples drawn from the product of marginals $\mathbb{P}_{\mathbf{M}}\mathbb{P}_{\mathbf{Y}^n}$. We use MINE for short blocklengths due to its high variance in high dimensions, and validate the results with other lower-bound estimators such as NWJ and SMILE~\cite{belghazi2018mutual, song2019understanding}.  {In contrast, $f$-DIME extends the variational framework to general $f$-divergences and uses derangement sampling, yielding more accurate and lower-variance MI estimates via
\begin{align}
&T_{\text{opt}} \!=\! \arg \max_{T}\E_{\mathbb{P}_{\mathbf{M}\mathbf{Z}^n}}\left[T(\mathbf{m},\mathbf{z}^n) \!-\! f^{*}\left(T(\mathbf{m}, \varsigma(\mathbf{z}^n))\right)\right], \\ 
  & {L}_{\text{eve}} \approx  \E_{\mathbb{P}_{\mathbf{M}\mathbf{Z}^n}}\left[\log \left((f^{*})^{\prime}(T_{\text{opt}}(\mathbf{m},\mathbf{z}^n))\right)\right] 
\end{align}
where $f^{*}$ is the Fenchel conjugate of $f:\R_{+}\rightarrow\R$ with $f(1)=0$, and $(f^{*})^{\prime}$ is its derivative. $T: \F_2^k\times\R^n \rightarrow \R$ is a neural network, and $\varsigma$ denotes a derangement that satisfies
\begin{align}
\mathbb{P}_{\varsigma(\mathbf{Z}^n)}(\varsigma(\mathbf{z}^n)|\mathbf{m}) = \mathbb{P}_{\mathbf{Z}^n}(\mathbf{z}^n).
\end{align}
We select the maximum estimate among the $f$-DIME variants based on the KL divergence, squared Hellinger distance, and generative adversarial network divergence\footnote{CLUB~\cite{cheng2020club} was also tested as an upper bound estimator but was found to be inaccurate, often exceeding the entropy $H(\mathbf{M})=3$.}.} We denote $\hat{I}_{\text{bob}}$ and $\hat{L}_{\text{eve}}$ as estimates computed using empirical expectations over $10^5$ samples.

\subsection{Security-Reliability Loss Function}
{We also incorporate the information leakage ${L}_{\text{eve}}$ into the loss function by formulating a constrained optimization problem as
\begin{equation}
\min_{(e_r, d_r)} \; J_{\text{CCE}} \quad \textrm{s.t.} \; {L}_{\text{eve}} \leq \tau
\end{equation}
where $\tau$ is the target leakage threshold in bits. To address this, we introduce the trade-off loss as
\begin{equation}
    J_\beta = J_{\text{CCE}} + \beta ({L}_{\text{eve}}-\tau)
\label{eq:target_leakage}
\end{equation}
with $\beta$ updated as 
\begin{align}
\beta \leftarrow [\beta + \eta ({L}_{\text{eve}}-\tau)]^+
\end{align}
starting from $\beta=0$ and learning rate $\eta = 0.01$, which may be reduced if significant oscillations occur in the leakage estimate. The training consists of two interleaved phases. We initialize the model $(e_r, d_r)$ with weights pre-trained using only the CCE loss. In Phase 1,  the $f$-DIME is retrained every $50$ batches to estimate the current information leakage. In Phase 2, $(e_r, d_r)$ is updated using the trade-off loss $J_\beta$.}

\section{RD Feedback Wiretap Code Simulation Results}

We next evaluate the seeded modular learned feedback codes in terms of security (information leakage) and reliability (BLER). We begin with the non-fading case ($H_Y = H_Z = 1$), followed by the fading case. The encoder and decoder are jointly trained to minimize the CCE loss, and we further study the trade-off between reliability and information leakage by incorporating leakage into the loss. We observe that leakage depends on the random seed, so we fix $\mathbf{S}$ as shared randomness between legitimate parties. During training, Bob and Eve share the same forward SNR, while during testing, Bob's SNR is fixed and Eve's SNR is varied\footnote{Simulations are implemented in PyTorch on an NVIDIA GeForce RTX 5090. The code will be made publicly available upon acceptance.}. Specifically, we train the WTC-Lightcode using the AdamW optimizer with an initial learning rate of $10^{-3}$ and a LambdaLR scheduler over $120$ epochs ($1000$ batches/epoch, $B=10^5$), while mutual information estimation employs a batch size of $5000$ over $300$ epochs.

\subsection{Secrecy Rate Effects}
\begin{figure}[!t]
    \centering
\includegraphics[width=1\columnwidth]{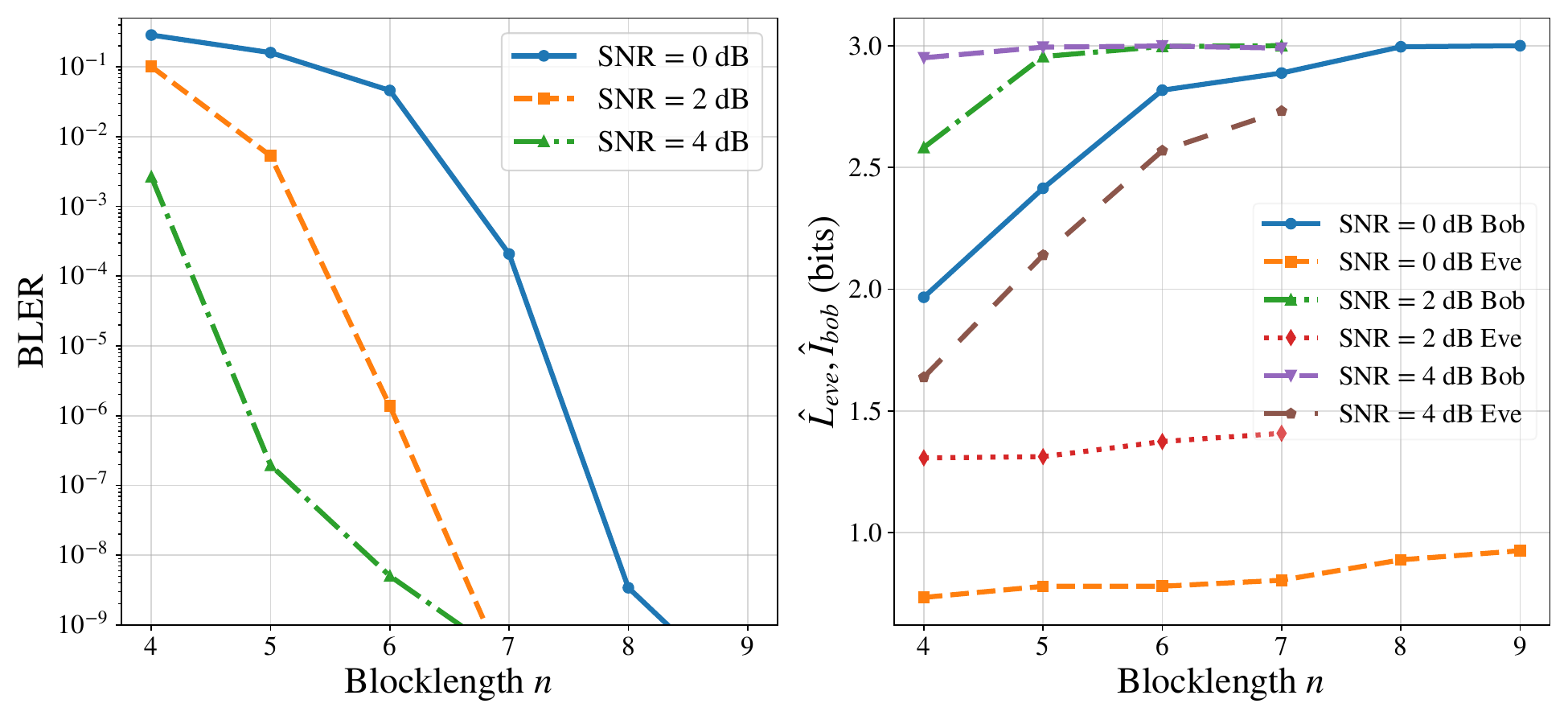}
\Description{We evaluate the WTC-Lightcode with $k =q= 3$ and noiseless feedback at forward SNR values of $0, 2, 4$ dB. The left plot shows that BLER decreases as blocklength $n$ increases. The right plot compares Bob's information ($\hat{I}_\text{bob}$) and Eve's leakage ($\hat{L}_\text{eve}$), showing that $\hat{I}_\text{bob}$ approaches 3 bits while the information leakage remains significantly lower across all blocklengths.}
    \caption{BLER (left) and estimated $\hat{I}_\text{bob}$, $\hat{L}_\text{eve}$ (right) versus blocklength under $\text{SNR} = \mathrm{\bar{S}}_{Y,f} = \mathrm{\bar{S}}_{Z,f}$ with noiseless feedback.}
    \label{fig:block_len}
\end{figure}

{We consider the baseline setting with noiseless feedback, CCE loss, and message length $k=3$. The random seeds $\mathbf{s}=[1, 0, 1]$, $[1, 1, 0, 1]$, and $[1, 1, 0, 1, 0]$ are fixed for $q=3,4,5$ and shared between the legitimate parties. As shown in Figure~\ref{fig:block_len}, we first study the effect of blocklength with $q=3$ (no random bits $\mathbf{B}$) and equal SNR ($\mathrm{\bar{S}}_{Y,f} = \mathrm{\bar{S}}_{Z,f}$) on reliability and security in WTC-Lightcode. Each additional round of error correction improves reliability, with BLER decreasing quickly and $\hat{I}_{\text{bob}}$ approaching $H(\mathbf{M}) = 3$ bits as Bob's observations nearly reveal the complete message. In contrast, the information leakage remains relatively small at low SNRs (0 and 2 dB) and increases only slightly with the blocklength $n$. At higher SNR (4 dB), leakage grows with blocklength, as Eve's reliable observations also influence the encoder output. This effect can be mitigated by adding random bits or imposing a leakage constraint. Notably, when Bob and Eve experience the same forward SNR, $\hat{I}_\text{bob}$ typically exceeds $\hat{L}_\text{eve}$, even in the RD case. This suggests that feedback creates additional shared randomness between Alice and Bob, providing feedback lunch, as defined above. We measure this effect by the \textit{security-advantage gain (SAG)} defined below.
\begin{definition}[SAG]\label{def:gain}
Given ${I}_{\text{bob}}$ at forward SNR $\mathrm{\bar{S}}_{Y,f}$, let $\mathrm{\bar{S}}_{Z,f}$ be the smallest SNR for Eve such that the leakage satisfies $|I_{\text{bob}}(\mathrm{\bar{S}}_{Y,f})- L_{\text{eve}}(\mathrm{\bar{S}}_{Z,f})| < \varepsilon$ for a small threshold $\varepsilon > 0$. The SAG is defined as
\begin{equation}
\Lambda_{(k, q, n, \mathrm{\bar{S}}_{Y,f})} \coloneqq \frac{1}{2} \log_2\frac{1 + \mathrm{\bar{S}}_{Z,f}}{1 + \mathrm{\bar{S}}_{Y,f}}.
\end{equation}
SAG roughly quantifies Eve's extra effort to match Bob's information level. Beyond this threshold, feedback offers no advantage, making secure communication impossible.
\end{definition}

\begin{remark}
    \normalfont The additional effort required for Eve to match the information level at Bob for feedback wiretap channels actually depends on the code rate, forward SNRs, feedback SNRs, and the length of the random sequence $q-k$ because of the finite blocklengths considered. However, SAG serves as a practical measure of the feedback lunch, which is more accurate than existing practical metrics in the literature, such as the metric that measures the SNR differences. Thus, we consider SAG for our analyses.
\end{remark}
}

\begin{table}[!t]
\centering
\caption{Effects of random bit sequence with $k=3$.}
\begin{tabular}{ccccccc}
\toprule
$\mathrm{\bar{S}}_{Y,f}$ (dB) & $q$ & $n$ & BLER & $\hat{I}_{\text{bob}}$ & $\hat{L}_{\text{eve}}$ & $\Lambda$ \\
\midrule
$0$  & $4$   & $9$ & $1.35\times 10^{-4}$      &       $2.97$                 &    $0.34$                   &     $4.48$\\
$0$ & $5$   & $9$ & $0.16$      &        $2.23$                &          $0.14$              &   $3.99$\\
$2$  & $4$   & $7$ & $6.91\times 10^{-4}$   &  $2.97$                   &        $1.14$              &      $4.96$   \\ 
$2$  & $5$   & $7$ & $0.11$   &  $2.33$                   &        $0.43$              &  $4.13$       \\ 
$4$  &  $4$  & $6$ & $5.25\times 10^{-5}$   &  $2.99$                   &    $1.69$                & $2.75$      \\ 
$4$  &  $5$  & $6$ & $2.24\times 10^{-2}$   &  $2.64$                   &    $0.80$                & $2.42$      \\ 
\bottomrule
\end{tabular}
\label{tab:random_bits}
\end{table}

We observe that the baseline SAG values are induced by channel-output feedback, specifically:
\begin{equation}
\Lambda_{(3,3,9,0)} = 0.89, \quad \Lambda_{(3,3,7,2)} = 1.41, \quad \Lambda_{(3,3,6,4)} = 0.67.
\end{equation}
Next, we study how the random bit length $q-k$ affects performance, with $\hat{I}_{\text{bob}}$ and $\hat{L}_{\text{eve}}$ evaluated at $\mathrm{\bar{S}}_{Y,f}$ (Table~\ref{tab:random_bits}). For fixed $n$, adding random bits reduces leakage. However, longer random sequences do not always help, as maximizing SAG requires both low BLER (or higher ${I}_{\text{bob}}$) and low leakage (${L}_{\text{eve}}$). Thus, selecting $q-k$ to maximize SAG at a given SNR is important to improve overall performance. For instance, we observe that $\Lambda_{(3,4,9, 0)} = 4.48$ exceeds $\Lambda_{(3,5,9, 0)} = 3.99$.

\subsection{Comparisons with Existing Feedback Codes}
\begin{figure}[!t]
    \centering
\includegraphics[width=0.99\columnwidth]{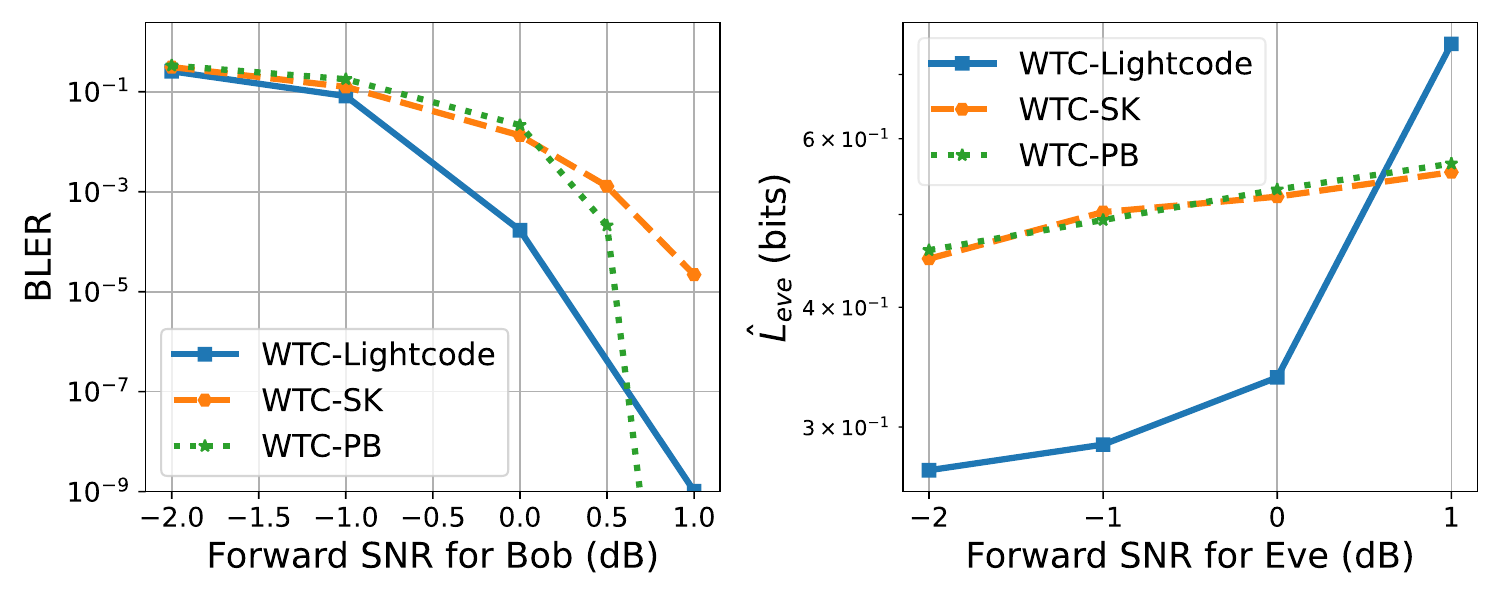}
\Description{Performance comparisons among WTC-Lightcode, WTC-SK, and WTC-PB show that WTC-Lightcode achieves the lowest BLER at low SNRs, while WTC-PB excels at high SNRs (e.g., 1 dB). The right plot illustrates that while information leakage increases with Eve's forward SNR, WTC-Lightcode maintains superior security at lower SNRs compared to other schemes.}
    \caption{BLER versus forward SNR at Bob (left) and $\hat{L}_\text{eve}$ versus forward SNR at Eve (right). Each data point is obtained under the condition $\mathrm{\bar{S}}_{Y,f}=\mathrm{\bar{S}}_{Z,f}$.}
    \label{fig:performance}
\end{figure}

We next compare the proposed WTC-Lightcode with other feedback codes, namely WTC-SK and WTC-PB that also exhibit a feedback lunch property. For example, we observe $\Lambda_{(3, 4, 9, 0)} = 4.15$ for WTC-SK, slightly below WTC-Lightcode ($4.48$). We set $q - k = 1$ with message length $k = 3$ and blocklength $n = 9$, which leverages the randomness introduced by the security layer while maintaining high BLER performance. We depict reliability and information leakage performance in Figure~\ref{fig:performance}. We observe that {WTC-Lightcode performs better at low SNRs in both BLER and information leakage, while WTC-PB outperforms WTC-Lightcode in BLER at high SNRs, achieves lower leakage than WTC-Lightcode, and is thus the preferred choice in that regime.}
{Moreover, WTC-SK exhibits worse BLER performance than WTC-Lightcode, while achieving similar information leakage to WTC-PB. This follows because WTC-SK and WTC-PB are identical during the first $n-1$ rounds, and in the final round, both transmit only the estimated error from previous rounds instead of the message $\mathbf{M}$. In contrast, WTC-Lightcode is a nonlinear scheme that incorporates feedback from Eve, resulting in different information leakage.} 

\subsection{Noisy Feedback Effects}
Codes that linearly encode the feedback into future transmissions, such as WTC-SK and WTC-PB, perform relatively poorly when the feedback is noisy, motivating the development of the nonlinear learned WTC-Lightcode, which is more robust to feedback noise. Table~\ref{tab:noisyfb} considers a message length of $k=3$, with $\hat{L}_{\text{eve}}$ evaluated under equal forward ($\mathrm{\bar{S}}_{Y,f} = \mathrm{\bar{S}}_{Z,f}$) and feedback ($\text{S}_{Y,fb}= \text{S}_{Z,fb}$) SNRs. We test both low and high SNR regimes, with and without random bits, where WTC-Lightcode maintains good BLER performance. We observe that noisier feedback channels increasingly disrupt the shared randomness between legitimate parties, leading to greater information leakage. However, the legitimate parties still achieve a positive SAG even under noisy feedback.

\begin{table}[!t]
\centering
\caption{Reliability and security with noisy feedback and $k=3$.}
\begin{tabular}{cccccccc}
\toprule
$\mathrm{\bar{S}}_{Y,f}$ & $\mathrm{S}_{Y,fb}$ & $q$ & $n$ & BLER & $\hat{I}_{\text{bob}}$ & $\hat{L}_{\text{eve}}$ & $\Lambda$ \\
\midrule
$0$ & $20$ & $3$ & $8$ & $2.21\times 10^{-3}$ & $2.97$ & $1.34$ & $2.50$ \\
$0$ & $10$ & $3$ & $8$ & $1.82\times 10^{-2}$ & $2.91$ & $2.10$ & $0.53$ \\
$0$ & $5$  & $3$ & $8$ & $3.8\times 10^{-2}$  & $2.83$ & $2.32$ & $0.29$ \\
$4$ & $20$ & $3$ & $6$ & $4.02\times 10^{-6}$ & $2.99$ & $2.56$ & $0.82$ \\
$4$ & $10$ & $3$ & $6$ & $8.44\times 10^{-4}$ & $2.99$ & $2.81$ & $0.53$ \\
$4$ & $5$  & $3$ & $6$ & $4.16\times 10^{-3}$ & $2.98$ & $2.88$ & $0.12$ \\
$1$ & $20$ & $4$ & $9$ & $3\times 10^{-3}$    & $2.96$ & $1.07$ & $3.73$ \\
$1$ & $10$ & $4$ & $9$ & $2.8\times 10^{-2}$  & $2.82$ & $2.17$ & $0.44$ \\
$1$ & $5$  & $4$ & $9$ & $4.1\times 10^{-2}$  & $2.78$ & $2.51$ & $0.09$ \\
\bottomrule
\end{tabular}
\label{tab:noisyfb}
\end{table}

\subsection{Security-Reliability Loss-Function Effects}
\begin{figure}[!t]
  \centering
  \includegraphics[width=\linewidth]{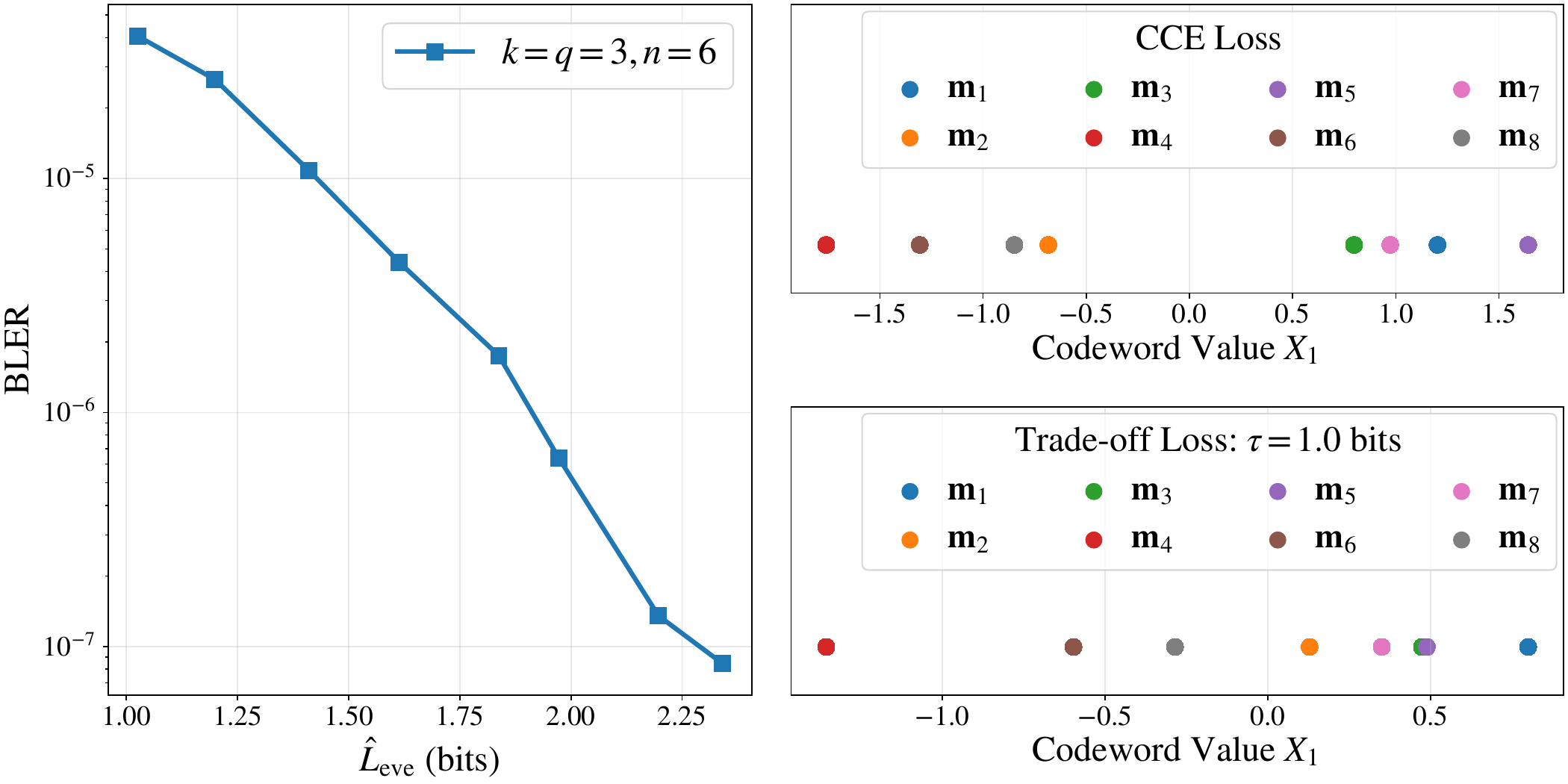}
  \Description{We evaluate the trade-off loss performance of the WTC-Lightcode and observe a clear trade-off between reliability and security. We also find that the encoder overlaps the generated codewords to reduce information leakage, making it difficult for Eve to distinguish between messages, but this comes at the cost of increased BLER.}
  \caption{BLER versus information leakage with forward SNR $\mathrm{\bar{S}}_{Y,f}=\mathrm{\bar{S}}_{Z,f} = 4$ dB (left) and first-round codewords $X_1$ (right).}
  \label{fig:control_leakage}
\end{figure}
For WTC-Lightcode, we observe that information leakage increases with the forward SNR, and the security layer alone only provides limited control over leakage. To achieve more flexible control, we introduce a trade-off loss that explicitly constrains leakage while preserving reliability. By specifying a target leakage $\tau$ in \eqref{eq:target_leakage}, the model is trained to approach this value. To test this, consider a case with $k=q=3$, $n=6$, $\mathrm{\bar{S}}_{Y,f}=\mathrm{\bar{S}}_{Z,f} = 4$ dB and noiseless feedback. Starting from a baseline leakage of $2.57$ bits (produced by standard CCE loss), we vary $\tau$ from $1.0$ to $2.4$ bits. As shown in Figure~\ref{fig:control_leakage} (left),  tuning $\tau$ effectively controls the leakage, showing a clear trade-off between the BLER and leakage. We next illustrate how the proposed trade-off loss affects reliability and information leakage. Since $\mathbf{M} \in \mathbb{F}_2^3$, it has $2^3$ realizations $\{\mathbf{m}_1,\ldots,\mathbf{m}_8\}$. Without feedback in the first round, the encoder output is $X_1 = e_r(\mathbf{M})$, assigning one real-valued codeword to each realization. Figure~\ref{fig:control_leakage} (right) shows the resulting $X_1$ values. Under the CCE loss, the points are well separated. 
Imposing, e.g., $\tau = 1.0$ bit causes the codewords corresponding to $\mathbf{m}_3$ and $\mathbf{m}_5$ to overlap, making them indistinguishable to Eve and reducing leakage at the cost of increased BLER.

\subsection{Fading Extensions for RD-WTC-F}
\begin{figure}[!t]
    \centering
\includegraphics[width=0.99\columnwidth]{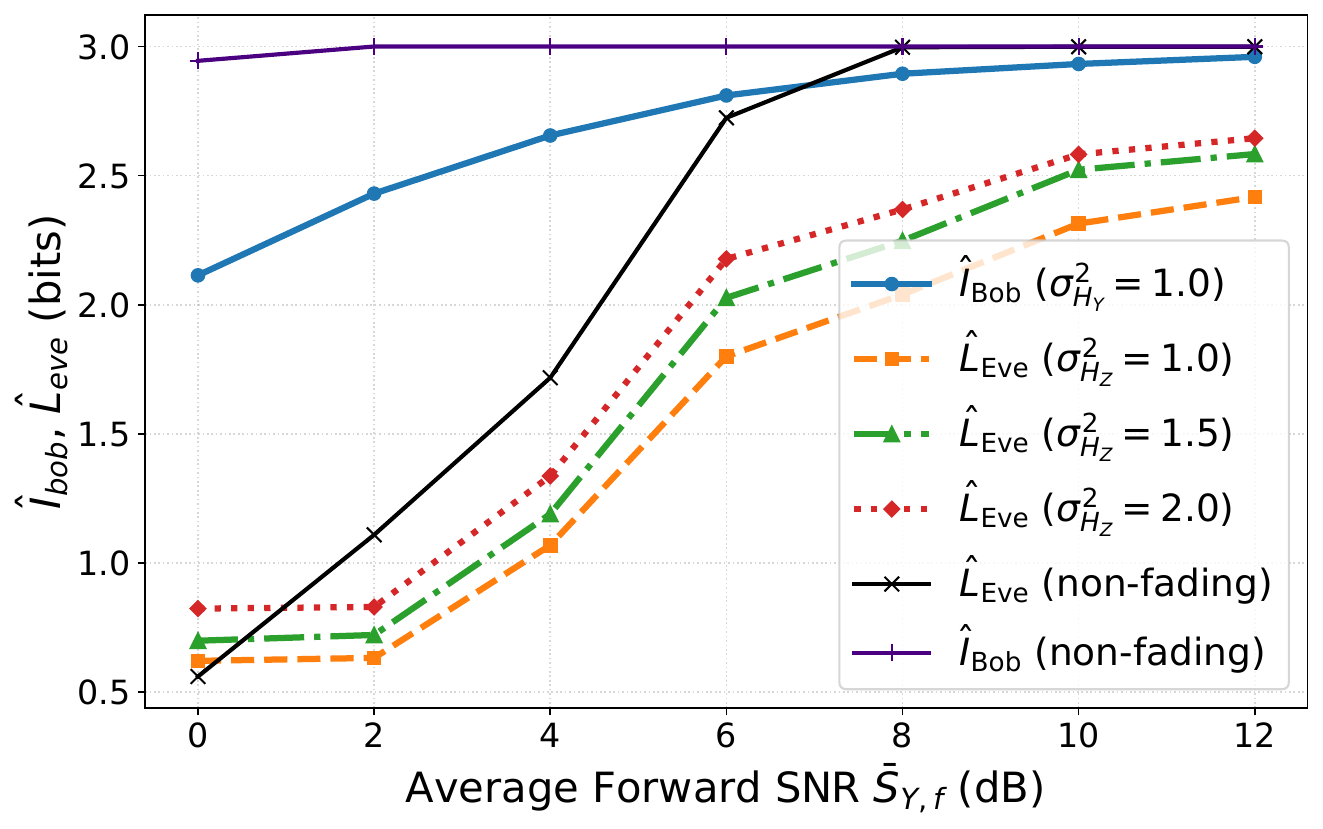}
\Description{We evaluate the fading WTC-Lightcode with $k =3$, $q = 4$, $n = 9$ under noiseless feedback. The plot shows Bob's information ($\hat{I}_\text{bob}$) and Eve's leakage ($\hat{L}_\text{eve}$) for different fading coefficients. As SNR increases, Bob's information approaches 3 bits while Eve's leakage remains controlled, even in RD cases.
}
    \caption{Estimated $\hat{I}_\text{bob}$ and $\hat{L}_\text{eve}$ versus the average forward SNR for Bob $\bar{S}_{Y,f}$ in the RD cases.}
    \label{fig:fading}
\end{figure}

In this section, we study feedback's impact on block-fading wiretap channels in the RD case. The information leakage $L_{\text{eve}}$ in the fading case has been defined as $I(\mathbf{M}; \mathbf{Z}^n\mid H_Z)$, which is equal to $I(\mathbf{M}; \mathbf{Z}^n, H_Z)$ as $\mathbf{M}$ and $H_Z$ are independent. Accordingly, we approximate $L_{\text{eve}}$ using a mutual information estimator with inputs $\mathbf{M}$ and the concatenated vector $[\mathbf{Z}^n, H_Z]$. The same procedure is used to estimate $I_{\mathrm{bob}}$. All estimations are computed using $10^5$ independent fading blocks of length $n$. As shown in Figure~\ref{fig:fading}, we study the case with $k=3$, $q = 4$, $n = 9$, and noiseless feedback over the fading wiretap channel, using the non-fading case as the baseline. By setting $\sigma_{N_Z}^2 = \sigma_{N_Y}^2$, we consider three ratios $\sigma_{H_Z}^2 /\sigma_{H_Y}^2 = 1, 1.5,$ and $2.0$, where a larger ratio corresponds to a better channel for Eve. In these scenarios, Eve achieves SNR advantages of $0$ dB, $1.76$ dB, and $3.01$ dB over Bob, respectively. Overall, the information leakage increases with the forward SNR $\mathrm{\bar{S}}_{Y,f}$. Compared with the fading cases, the leakage in the non-fading channel grows more sharply. Interestingly, at a low forward SNR of $\mathrm{\bar{S}}_{Y,f} = 0$ dB, the fading cases exhibit higher leakage than the non-fading case. However, as the forward SNR increases, the fading effect becomes beneficial for secrecy, resulting in lower leakage than the non-fading baseline. Although a better channel condition for Eve (larger $\sigma_{H_Z}^2$) leads to higher leakage, the overall information leakage remains small relative to $\hat{I}_\text{bob}$, even in these RD scenarios. These results show that feedback enhances secrecy in fading wiretap channels.

\section{Conclusion}
We studied a seeded modular design, termed WTC-Lightcode, for the Gaussian WTC-F, combining 2-UHF for security and learned codes for reliability. With (noisy) feedback, even in the RD case, legitimate parties have a SAG over Eve thanks to the feedback lunch. We illustrated the effects of different parameters and compared the performance of various feedback code designs. Furthermore, we incorporated information leakage into the training loss to enable a flexible trade-off between BLER and leakage. Finally, the scheme extends to fading channels, where a SAG over Eve is also observed. Our results motivate new feedback code designs for secure ISAC applications in sixth-generation systems, where channel-output feedback is intrinsic. In future, we will also adapt hybrid code constructions, such as in \cite{OnurConcatenated}, to enable larger blocklengths.

\begin{acks}
N. Devroye and Y. Zhou were partially supported by NSF award 2240532. The contents of this article are solely the responsibility of the authors and do not necessarily represent the official views of the NSF. O. G\"{u}nl\"{u}'s work was partially supported by the German Federal Ministry of Research, Technology and Space (BMFTR) 6GEM+ Transfer Hub under Grants 16KIS2412 and 16KISS005.
\end{acks}

\bibliographystyle{ACM-Reference-Format}
\bibliography{reference}








\end{document}